\documentstyle[prl,aps,multicol]{revtex}
\begin{document}
\draft
\title{Parity Effect in Ground State Energies of Ultrasmall
  Superconducting Grains} 

\author{K.A. Matveev$^{1,2}$ and A.I. Larkin$^{3,4}$} 

\address{$^{1}$Duke University, Department of Physics, 
               Durham, NC 27708-0305\\ 
         $^{2}$Massachusetts Institute of Technology,
               Cambridge, MA 02139\\
         $^{3}$Theoretical Physics Institute, University of Minnesota, 
               Minneapolis MN 55455\\
         $^{4}$L.D. Landau Institute for Theoretical Physics, 
               117940 Moscow, Russia} 
\date{January 7, 1997}
\maketitle

\begin{abstract}
  We study the superconductivity in small grains in the regime when the
  quantum level spacing $\delta\varepsilon$ is comparable to the gap
  $\Delta$. As $\delta\varepsilon$ is increased, the system crosses over
  from superconducting to normal state.  This crossover is studied by
  calculating the dependence of the ground state energy of a grain on the
  parity of the number of electrons.  The states with odd numbers of
  particles carry an additional energy $\Delta_P$, which shows
  non-monotonic dependence on $\delta\varepsilon$. Our predictions can be
  tested experimentally by studying the parity-induced alternation of
  Coulomb blockade peak spacings in grains of different sizes.

\end{abstract}


\begin{multicols}{2}
  
The standard BCS theory\cite{BCS} gives a good description of the
phenomenon of superconductivity in large samples. However, it was noticed
by Anderson\cite{Anderson} in 1959 that as the size of a superconductor
becomes smaller, and the quantum level spacing in the sample
$\delta\varepsilon$ approaches the superconducting gap $\Delta$, the BCS
theory fails. The interest to the superconductivity in such {\it
  ultrasmall\/} grains was renewed by recent experiments by Ralph, Black
and Tinkham\cite{Ralph,Black}, who fabricated and studied nanometer-scale
aluminum grains. In qualitative agreement with the prediction
\cite{Anderson}, they demonstrated \cite{Black} the existence of
superconducting gap in relatively large grains, with estimated level
spacings $\delta\varepsilon \approx 0.02$ and 0.08 meV smaller than the
superconducting gap $\Delta\approx 0.31$ meV, whereas no signs of
superconductivity were observed \cite{Ralph} in smaller grains,
$\delta\varepsilon \approx 0.7$ meV. These experiments raise a theoretical
question about the nature of the crossover from superconducting to normal
state in ultrasmall particles with level spacings $\delta\varepsilon \sim
\Delta$.

This problem was addressed in two recent theoretical papers.  J. von Delft
{\it et al.}\cite{vonDelft} explored the BCS gap equation in a finite-size
system with equidistant discrete energy levels, and found that as the
level spacing is increased, the superconducting gap of the grain vanishes
at a certain critical value of $\delta\varepsilon$, which is of order
$\Delta$ and depends on the parity of the total number of electrons in the
grain.  Smith and Ambegaokar\cite{Smith} extended the treatment of
Ref.~\cite{vonDelft} to take into account Wigner-Dyson fluctuations of the
energy levels in the grain.

It is worth noting that the theories \cite{vonDelft,Smith} treat the
superconductivity in small grains within the selfconsistent mean field
approximation for the superconducting order parameter. Although this
approximation works well for large systems, one should expect the quantum
fluctuations of the order parameter to grow when the level spacing
$\delta\varepsilon$ reaches $\Delta$. In this paper we present a theory of
superconductivity in ultrasmall grains which includes the effects of
quantum fluctuations of the order parameter.  We show that the corrections
to the mean field results which are small in large grains,
$\delta\varepsilon\ll \Delta$, become important in the opposite limit,
$\delta\varepsilon\gg\Delta$.

The superconducting gap $\Delta$ studied in Refs.~\cite{vonDelft,Smith} is
not well defined in the presence of quantum fluctuations. Therefore, we
must first identify an {\it observable\/} physical quantity which
characterizes the superconducting properties of small grains. The most
convenient such quantity for our purposes is the ground state energy of
the grain $E_N$ as a function of the number of electrons $N$.  More
precisely, we study the so-called {\it parity effect\/} in ultrasmall
grains, which is described quantitatively by parameter
\begin{equation}
  \label{Delta_P}
  \Delta_P = E_{2l+1} - \frac12\big(E_{2l}+E_{2l+2}\big).
\end{equation}
In the ground state of a large superconducting grain with an odd number of
electrons, one electron is unpaired and carries an additional energy
$\Delta_P=\Delta$. This result is well known in nuclear physics and was
recently discussed in connection to superconducting grains in
Refs.~\cite{Janko,Golubev}.  The parity effect was demonstrated
experimentally in Refs.~\cite{Lafarge,Tuominen}, where the Coulomb
blockade phenomenon\cite{blockade} in a superconducting grain was studied.
In a Coulomb blockade experiment one can measure $\Delta_P$ explicitly as
the difference of spacings between three neighboring peaks of linear
conductance through a superconducting grain.

We describe the grain by the following Hamiltonian:
\begin{equation}
  \label{Hamiltonian}
  \hat H 
      = \sum_{k\sigma} \varepsilon_k^{} a_{k\sigma}^\dagger a_{k\sigma}^{}
      - g \sum_{kk'} a_{k\uparrow}^\dagger a_{k\downarrow}^\dagger
                     a_{k'\downarrow}^{}a_{k'\uparrow}^{}.
\end{equation}
Here $k$ is an integer numbering the single particle energy levels
$\varepsilon_k$, the average level spacing $\langle \varepsilon_{k+1} -
\varepsilon_k \rangle =\delta\varepsilon$, operator $a_{k\sigma}$
annihilates an electron in state $k$ with spin $\sigma$, and $g$ is the
interaction constant.  In Eq.~(\ref{Hamiltonian}) we assume zero magnetic
field, so that the electron states can be chosen to be invariant under the
time reversal transformation\cite{Anderson}. We include in
Eq.~(\ref{Hamiltonian}) only the matrix elements of the interaction
Hamiltonian responsible for the superconductivity; the contributions of
the other terms are negligible in the weak coupling regime
$g/\delta\varepsilon\ll1$ we consider. Finally, we did not include in
Eq.~(\ref{Hamiltonian}) the charging energy responsible for the Coulomb
blockade, as its contribution to the ground-state energy is trivial.

In the absence of interactions, $g=0$, the parity parameter $\Delta_P$ can
be easily calculated. Indeed, the ground state energy $E_N$ is found by
summing up $N$ lowest single-particle energy levels.  This results in
$E_{2l+1}=E_{2l}+\varepsilon_{l+1}$ and
$E_{2l+2}=E_{2l}+2\varepsilon_{l+1}$. Substituting this into
Eq.~(\ref{Delta_P}), we find that without the interactions
$\Delta_P=0$. 

For weak interactions one can start with the first-order perturbation
theory in $g$. In this approximation an electron in state $k$ interacts
only with an electron with the opposite spin in the same orbital state
$k$.  Thus when the ``odd'' \mbox{$(2l+1)$-st} electron is added to the
grain, it is the only electron in the state $l+1$ and does not contribute
to the interaction energy, $\delta E_{2l+1}=\delta E_{2l}$. The next,
\mbox{$(2l+2)$-nd} electron goes to the same orbital state and interacts
with it: $\delta E_{2l+2}=\delta E_{2l+1}-g$. From Eq.~(\ref{Delta_P}) we
now find
\begin{equation}
  \label{Delta_P-first-order}
  \Delta_P = \frac{g}{2}, \quad {\rm at\ }g\to0.
\end{equation}

One should note that the result (\ref{Delta_P-first-order}) is not quite
satisfactory even in the weak coupling case $g/\delta\varepsilon\ll1$.
Indeed, the low-energy properties of a superconductor are usually
completely described by the gap $\Delta$. The interaction constant $g$ is
related to the gap $\Delta$ in a way which depends on a particular
microscopic model, so the result (\ref{Delta_P-first-order}) cannot be
directly compared with experiments.

This problem can be resolved by considering corrections of higher orders
in $g$, which are known\cite{AGD} to give rise to logarithmic
renormalizations of $g$. In the leading-logarithm approximation the
renormalized interaction constant is found \cite{AGD} as
\begin{equation}
  \label{g-renormalized}
  \tilde g = \frac{g}{1-\frac{g}{\delta\varepsilon}\ln\frac{D_0}{D}}. 
\end{equation}
Here $D_0$ is the high-energy cutoff of our model, which has the physical
meaning of Debye frequency, and $D\ll D_0$ is the low-energy cutoff. At
zero temperature, $D\sim\delta\varepsilon$. Taking into account the
relation between the gap in a large grain $\Delta$ and microscopic
interaction constant, $\Delta\sim D_0 e^{-\delta\varepsilon/g}$, we find
with logarithmic accuracy $\tilde g =
\delta\varepsilon/\ln(\delta\varepsilon/\Delta)$.  Finally, substituting
the renormalized interaction constant into
Eq.~(\ref{Delta_P-first-order}), we get
\begin{equation}
  \label{Delta_P-renormalized}
  \Delta_P =
     \frac{\delta\varepsilon}{2\ln\frac{\delta\varepsilon}{\Delta}},
  \quad \Delta\ll\delta\varepsilon. 
\end{equation}
Unlike the first-order result (\ref{Delta_P-first-order}), $\Delta_P$ is
now expressed in terms of experimentally observable parameters $\Delta$
and $\delta\varepsilon$ rather than model-dependent interaction constant
$g$.

It is instructive to compare Eq.~(\ref{Delta_P-renormalized}) with the
results of Refs.~\cite{vonDelft,Smith}. In a very small grain with
$\delta\varepsilon\gg\Delta$, the mean field gap studied in
Refs.~\cite{vonDelft,Smith} vanishes, and no parity effect is expected.
On the contrary, our result (\ref{Delta_P-renormalized}) predicts that in
small grains the parity effect is {\it stronger\/} than in the large ones.
This behavior is due to the strong quantum fluctuations of the order
parameter which persist even when its mean field value studied in
Refs.~\cite{vonDelft,Smith} vanishes. The physics of the fluctuations of
the order parameter is hidden in the renormalization procedure leading to
Eq.~(\ref{g-renormalized}). Below we present a different technique, which
explicitly shows the role of the fluctuations.  It will allow us to
rigorously derive Eq.~(\ref{Delta_P-renormalized}) and to study the
fluctuation corrections in the case of large grains,
$\delta\varepsilon\ll\Delta$.

A convenient way to treat the fluctuations of the order parameter is by
using path integral technique\cite{Ambegaokar}. This approach gives an
exact expression for the grand partition function of a superconductor:
\begin{equation}
  \label{grand}
  Z(\mu,T) = {\rm Tr} \exp\left(-\frac{\hat H-\mu \hat N}{T}\right),
  \quad \hat N = \sum_{k\sigma}a_{k\sigma}^\dagger a_{k\sigma}^{}.
\end{equation}
Here $\mu$ is the chemical potential, $T$ is the temperature, and $\hat N$
is the operator of the number of electrons in the grain. At $T\to0$ the
dominating term in $Z(\mu,T)$ corresponds to the ground state of the grain
with a certain number of electrons:
\begin{equation}
  \label{Omega}
  Z(\mu,T\to0)= e^{-\Omega(\mu)/T}, 
  \quad 
  \Omega(\mu) = \min_N\{E_N-\mu N\}.
\end{equation}
Thus we can find the ground state energy $E_N$ by studying the grand
partition function (\ref{grand}).

One problem with this method of calculating $E_N$ is that because of the
parity effect (\ref{Delta_P}) with $\Delta_P>0$ the odd charge states do
not contribute to $Z(\mu,T\to0)$. To find $E_{2l+1}$ let us consider the
effect of interactions on the unperturbed ground state of $2l+1$
electrons. Since the state $l+1$ is filled with one electron, the
interaction term in the Hamiltonian (\ref{Hamiltonian}) can neither create
nor destroy a pair in this state. Thus $E_{2l+1}$ can be found as
\begin{equation}
  \label{odd_recipe}
  E_{2l+1} = \varepsilon_{l+1} + \tilde E_{2l},
\end{equation}
where $\tilde E_{2l}$ is the ground state energy of a grain with $2l$
electrons for the system (\ref{Hamiltonian}) with state $k=l+1$ excluded.

The idea of the path integral approach\cite{Ambegaokar} is to replace the
formulation (\ref{Hamiltonian}) of the problem in terms of electronic
operators $a_{k\sigma}$ by an equivalent formulation in terms of the
superconducting order parameter $\Delta(\tau)$. The latter is introduced
as an auxiliary field for a Hubbard-Stratonovich transformation splitting
the quartic interaction term in Eq.~(\ref{Hamiltonian}) into quadratic
pair creation and annihilation operators. Then the trace over the
fermionic variables can be calculated, and one finds
\begin{equation}
  \label{path-integral}
  Z(\mu,T) = \int D^2\Delta(\tau) e^{-S[\Delta]},
\end{equation}
where the action $S[\Delta]$ is defined as
\begin{equation}
  \label{action}
  S[\Delta] = 
     -\sum_k \!\left[{\rm Tr} \ln \hat G^{-1}_k-\frac{\xi_k}{T}\right]\!
              + \frac{1}{g}\int_0^{1/T}\!\!|\Delta(\tau)|^2d\tau.
\end{equation}
Here $\xi_k=\varepsilon_k-\mu$, and the inverse Green's function
\begin{eqnarray}
  \hat G^{-1}_{k}(\tau,\tau') &=& 
     \left[-\frac{d}{d\tau} -\xi_k\sigma^z
      - \Delta(\tau)\sigma^{+} - \Delta^{*}(\tau)\sigma^{-}\right]
     \nonumber\\
  \label{Green}
     & & \times\delta(\tau-\tau'),
\end{eqnarray}
where $\sigma^{\pm} = \sigma^x \pm i\sigma^y$, and $\sigma^{x,y,z}$ are
the standard Pauli matrices. $\hat G^{-1}$ satisfies antiperiodic boundary
conditions: $\hat G^{-1}_{k}(\tau+T^{-1})=-\hat G^{-1}_{k}(\tau)$.

Unlike in the case of large superconductors\cite{Ambegaokar}, the order
parameter $\Delta$ in Eqs.~(\ref{path-integral})--(\ref{Green}) does not
depend on the coordinates, and thus the contributions of different states
$k$ in the action (\ref{action}) decouple. This results from the
simplified form of the interaction term in the Hamiltonian
(\ref{Hamiltonian}). The space fluctuations of $\Delta$ are negligible for
grains smaller than the coherence length of the superconductor; this
condition is well satisfied in ultrasmall grains. On the other hand, the
time fluctuations of $\Delta$ accounted for in
Eqs.~(\ref{path-integral})--(\ref{Green}) lead to the corrections to the
mean field BCS theory and are studied below.

First we consider the regime of weak interactions,
$\Delta\ll\delta\varepsilon$. In this case the $\Delta$-dependent terms
can be considered to be a small perturbation $\hat V=\Delta\sigma^+
+\Delta^*\sigma^-$, and one can formally expand the action (\ref{action})
in power series in $\hat V$ using
\begin{equation}
  \label{algebra}
  {\rm Tr}\ln(\hat G^{-1}_0 - \hat V) = {\rm Tr}\ln\hat G^{-1}_0
    -\sum_{j = 1}^{\infty} \frac{1}{j} {\rm Tr} (\hat G_0 \hat V)^j.
\end{equation}
The first-order term vanishes because matrix $\hat V$ is off-diagonal, so
we study the quadratic in $\Delta$ contribution to the action. The
calculations are more convenient to perform in terms of the Fourier
components $\Delta_m$ of the order parameter, defined in a usual way:
\begin{equation}
  \label{Delta_m}
  \Delta(\tau) = T \sum_m \Delta_m e^{-i\omega_m\tau},
  \quad \omega_m = 2\pi T m.
\end{equation}
The calculation of the second-order contribution to the action
(\ref{action}) is straightforward and gives
\begin{equation}
  \label{delta_S}
  \delta S = T\sum_m \frac{1-\alpha(i\omega_m)}{g}|\Delta_m|^2,
  \quad
  \alpha(E) = g\sum_k\frac{{\rm sgn\,} \xi_k}{2\xi_k - E}.
\end{equation}

The functional integral (\ref{path-integral}) is now easily evaluated by
integrating over the real and imaginary parts of each $\Delta_m$. We
normalize the result for the partition function $Z$ by its value $Z_0$
for non-interacting system, which corresponds to $\alpha=0$ in
Eq.~(\ref{delta_S}),
\begin{eqnarray}
  \frac{Z(\mu,T)}{Z_0(\mu,T)}
  &=& \prod_m \frac{1}{1-\alpha(i\omega_m)}
      \nonumber\\
  &=& \prod_m \prod_k \frac{2\xi_k -i\omega_m}{2\tilde\xi_k -i\omega_m}
      = \prod_k \frac{\sinh(\xi_k/T)}{\sinh(\tilde\xi_k/T)}.
  \label{Z_weak}
\end{eqnarray}
Here $\tilde\xi_k$ are defined by $1-\alpha(2\tilde\xi_k)=0$.  Assuming
weak interactions, $\Delta\ll\delta\varepsilon$, we find $\tilde\xi_k =
\xi_k + \delta\xi_k$, where
\begin{equation}
  \label{delta_xi}
  \delta\xi_k = -\frac{g}{2}\,
          \frac{{\rm sgn\,}\xi_k}
          {1-\frac{g}{2}\sum_{k'\neq k}\frac{{\rm sgn\,}\xi_{k'}}
                                            {\xi_{k'}-\xi_k}}.
\end{equation}
We can now compare the result (\ref{Z_weak}) with the definition
(\ref{Omega}) of $\Omega(\mu)$ and find
\begin{equation}
  \label{Omega_weak}
  \delta\Omega(\mu) = \sum_k \delta\xi_k \,{\rm sgn\,}\xi_k.
\end{equation}

One can easily see that for $\varepsilon_l<\mu<\varepsilon_{l+1}$ this
correction does not depend on $\mu$. According to (\ref{Omega}) such
$\delta\Omega$ should be interpreted as the correction $\delta E_{2l}$ to
the ground state energy of $2l$ electrons present in the grain in this
range of $\mu$. We then use the rule (\ref{odd_recipe}) to find $\delta
E_{2l+1}$. To do this we exclude the state $k=l+1$ from the sums in
Eq.~(\ref{delta_xi}) and (\ref{Omega_weak}) and calculate $\delta
E_{2l}'$. The result for $\Delta_P$ coincides with
Eq.~(\ref{Delta_P-renormalized}).

We now turn to the case of stronger interactions,
$\Delta\gg\delta\varepsilon$. A good starting point in this regime is the
standard BCS theory\cite{BCS}, which corresponds to a mean-field
approximation for the order parameter in the path integral
approach\cite{Ambegaokar}. Substituting a time-independent $\Delta$ into
the action (\ref{action}), one finds
\begin{equation}
  \label{mean-field}
  \Omega(\mu) = \sum_k(\xi_k - \epsilon_k) + \frac{1}{g}|\Delta|^2.
\end{equation}
Here $\epsilon_k = (\xi_k^2 + |\Delta|^2)^{1/2}$, and the value of
$|\Delta|$ must be chosen in a way which minimizes $\Omega$. This means
that $|\Delta|$ is determined from the usual BCS equation:
\begin{equation}
  \label{mean-field-Delta}
  \sum_k \frac{1}{2\epsilon_k} = \frac{1}{g}.
\end{equation}
In the continuous limit $\delta\varepsilon/\Delta\to 0$ one can apply the
rule (\ref{odd_recipe}) and find that the exclusion of one state from the
sum over $k$ in Eq.~(\ref{mean-field}) results in the energies of odd
charge states exceeding those of the even ones by $\Delta_P = \Delta$.

To find the corrections to $\Delta_P$ due to the finite
$\delta\varepsilon$ a more careful treatment of the mean-field
approximation is required. One can easily see that not only a
time-independent $\Delta_0(\tau)=|\Delta|$, but also any path
$\Delta_M(\tau)=|\Delta|e^{i2\pi MT\tau}$ with integer $M$ is a minimum of
the action (\ref{action}) which must be taken into account. It is
convenient to treat the path $\Delta_M(\tau)$ in Eq.~(\ref{Green}) by
performing a gauge transformation $\hat U = \exp(i\pi MT\tau\sigma^z)$,
which eliminates the time dependence of $\Delta_M(\tau)$ and shifts the
chemical potential $\mu\to\mu-i\pi MT$\cite{remark}. Thus instead of
$Z=e^{-\Omega/T}$ the partition function at $T\to0$ is now
\[
  Z(\mu,T)=\sum_{M} e^{-\Omega(\mu-i\pi MT)/T}
              =e^{-\Omega(\mu)/T}\sum_{M} e^{i\pi\Omega'(\mu)M},
\]
where $\Omega(\mu)$ is given by Eq.~(\ref{mean-field}). We have expanded
$\Omega(\mu-i\pi MT)$ in Taylor series in $i\pi M T$, and neglected the
terms vanishing at $T\to0$. It is now obvious that $Z(\mu,T)=0$ unless the
derivative of $\Omega$ is an even integer:
\begin{equation}
  \label{mu_2l}
  \Omega'(\mu_{2l}) = -2l.
\end{equation}
Thus the mean field approximation can be applied only for discrete values
of chemical potential $\mu_{2l}$ corresponding to solutions with $2l$
electrons. For odd number of electrons $\mu_{2l+1}$ is found as $\mu_{2l}$
in a system with one state $k=l+1$ at the Fermi level excluded. At
$\Delta\gg\delta\varepsilon$ one always gets $\mu_{N+1}-\mu_{N} =
\delta\varepsilon/2$.

To find $\Delta_P$ we substitute in Eq.~(\ref{Delta_P}) the ground state
energy as $E_N=\Omega(\mu_N) + \mu_N N$. The contribution of the second
term to $\Delta_P$ is $-\delta\varepsilon/2$. In evaluating the
contribution of $\Omega(\mu_N)$ one has to take into account the
dependence of $\mu_N$ on $N$ and the suppression\cite{Janko,Golubev} of
the self-consistent gap in Eq.~(\ref{mean-field}) $\Delta_{\rm
  odd}=\Delta-\delta\varepsilon/2$ for odd $N$ due to the exclusion of one
state $k$ from the gap equation (\ref{mean-field-Delta}). Combining all
the contributions, we get the following mean-field result:
\begin{equation}
  \label{Delta_P-mean-field}
  \Delta_P = \Delta - \frac{\delta\varepsilon}{2},
  \quad
  \delta\varepsilon\ll\Delta.
\end{equation}
It is interesting that a similar quantity $\tilde\Delta_P=-E_{2l} +
(E_{2l-1} + E_{2l+1})/2$ is unaffected by finite level spacing up to the
terms linear in $\delta\varepsilon$, i.e., $\tilde\Delta_P=\Delta$. Thus
at $\delta\varepsilon\ll\Delta$ we have $\Delta_P=\Delta_{\rm odd}$ and
$\tilde\Delta_P=\Delta_{\rm even}$.

Though the result $\Delta_P = \Delta$ can be obtained from the mean field
theory, an evaluation of corrections to it due to the level spacing
requires taking into account the effects of the fluctuations of the order
parameter. One can find the contribution of the fluctuations by expanding
the action near $\Delta(\tau)=|\Delta|$ using Eq.~(\ref{algebra}). The
second-order correction to the action is\cite{in_preparation}
\[
  \delta S = T
   \sum_m\left|\delta_{m}^{R}\gamma_{m}^{1/2} 
            +i\delta_{m}^{I}\gamma_{m}^{-1/2}\right|^2
   \sum_k \frac{\omega_m}{2\epsilon_k(\omega_m\gamma_m-2i\xi_k)},
\]
where $\delta_{m}^{R,I}$ are the Fourier components of the real and
imaginary parts of the fluctuation $\Delta(\tau) - |\Delta|$, and
$\gamma_m = (1+4|\Delta|^2/\omega_{m}^2)^{1/2}$. Now we evaluate the path
integral (\ref{path-integral}) by integrating over all $\delta_{m}^{R,I}$
and find the contribution of the fluctuations to $\Omega(\mu)$,
\[
  \delta\Omega(\mu) = \int_{-\infty}^{\infty} \frac{d\omega}{2\pi}
         \ln\left|\sum_k \frac{g|\omega|}{2\epsilon_k
         \big(\sqrt{\omega^2+4|\Delta|^2} -2i\xi_k\big)} \right|.
\]
A comparison of the expressions for $\delta\Omega$ in the cases of even
and odd numbers of electrons shows that they coincide up to the terms of
order $\delta\varepsilon$. Thus the fluctuations of the order parameter do
not affect the result (\ref{Delta_P-mean-field}).

Finally, we discuss the mesoscopic fluctuations of $\Delta_P$ given by our
results (\ref{Delta_P-renormalized}) and (\ref{Delta_P-mean-field}). It is
clear from Eq.~(\ref{delta_xi}) that unlike $\delta\varepsilon$ in the
numerator of Eq.~(\ref{Delta_P-renormalized}), the one in the argument of
the logarithm is sensitive to the Wigner-Dyson fluctuations of $\xi_k$.
Thus the relative mesoscopic fluctuation of the result
(\ref{Delta_P-renormalized}) is small,
$\delta\Delta_P/\Delta_P\sim1/\ln(\delta\varepsilon/\Delta)$. It is also
interesting to compare the mesoscopic fluctuation of the gap $\Delta$
originating from the level fluctuations in Eq.~(\ref{mean-field-Delta})
with the small corrections in Eq.~(\ref{Delta_P-mean-field}). One can
easily express\cite{in_preparation} the correction to $\Delta$ in terms of
the correction to the density of states $\nu(\xi)$ in
Eq.~(\ref{mean-field-Delta}). Then the mean-square fluctuation of the gap
is found using the well-known results for the correlator
$\langle\nu(\xi)\nu(\xi')\rangle$, and we get
$\sqrt{\langle(\delta\Delta)^2\rangle}=\delta\varepsilon/\pi\sqrt{2}$.

In conclusion, we have studied the parity effect (\ref{Delta_P}) in the
ground state energies of an ultrasmall superconducting grain. Although the
quantum fluctuations of the order parameter can be neglected for large
grains, Eq.~(\ref{Delta_P-mean-field}), they play a crucial role in small
grains, Eq.~(\ref{Delta_P-renormalized}). As the size of the grain
decreases, the parity effect first weakens,
Eq.~(\ref{Delta_P-mean-field}), but then starts increasing,
Eq.~(\ref{Delta_P-renormalized}). Thus we expect a minimum of $\Delta_P$
at a certain size of the grain such that $\delta\varepsilon\sim\Delta$.

\end{multicols}

\begin{references}
\bibitem{remark} For odd $M$ this transformation also changes the boundary
  conditions for $G^{-1}(\tau)$ from antiperiodic to periodic, which is
  irrelevant at $T\to0$.
\end{references}
\end{document}